\documentclass[%
preprint,
 amsmath,amssymb,
 aps,
prb,
]{revtex4-1}

\usepackage{graphicx}
\usepackage{float}
\usepackage{subfig}
\usepackage{caption}
\begin{document}

\title{Scanning transmission electron microscopy under controlled low-pressure atmospheres}

\author{Gregor T. Leuthner}
\author{Stefan Hummel}
\author{Clemens Mangler}
\author{Timothy J. Pennycook}
\author{Toma Susi}
\author{Jannik C. Meyer}
\author{Jani Kotakoski}
\email{jani.kotakoski@univie.ac.at}
\affiliation{University of Vienna, Faculty of Physics, Boltzmanngasse 5, 1090 Vienna, Austria}

\date{\today}

\begin{abstract}
Transmission electron microscopy (TEM) is carried out in vacuum to minimize the interaction of the imaging electrons with gas molecules while passing through the microscope column. Nevertheless, in typical devices, the pressure remains at $10^{-7}$~mbar or above, providing a large number of gas molecules for the electron beam to crack, which can lead to structural changes in the sample. Here, we describe experiments carried out in a modified scanning TEM (STEM) instrument, based on the Nion UltraSTEM~100. In this instrument, the base pressure at the sample is around $2\times 10^{-10}$~mbar, and can be varied up to $10^{-6}$~mbar through introduction of gases directly into the objective area while maintaining atomic resolution imaging conditions. We show that air leaked into the microscope column during the experiment is efficient in cleaning  graphene samples from contamination, but ineffective in damaging the pristine lattice. Our experiments also show that exposure to O$_2$ and H$_2$O lead to a similar result, oxygen providing an etching effect nearly twice as efficient as water, presumably due to the two O atoms per molecule. H$_2$ and N$_2$ environments have no influence on etching. These results show that the residual gas environment in typical TEM instruments can have a large influence on the observations, and show that chemical etching of carbon-based structures can be effectively carried out with oxygen.\end{abstract}


\maketitle

\section{\label{sec:intro}Introduction}

Aberration-corrected~\cite{Krivanek08UM} (S)TEM is an indispensable tool to study the atomic structure of novel materials such as nanotubes~\cite{Suenaga07NN,Sadan08PNAS}, nanoparticles~\cite{Hansen02S}, graphene~\cite{Meyer08NL} and other two-dimensional materials~\cite{jin_fabrication_2009,meyer_selective_2009,brivio_ripples_2011}. However, since energetic electrons must traverse the sample in order to scatter from its constituent atoms to provide an image, structural changes can take place in the sample during imaging~\cite{Egerton13UM}. 
Understanding any atomic-scale processes that occur during imaging is of fundamental importance for the correct interpretation of the recorded images. The most important causes for changes in the atomic structure under electron irradiation are knock-on damage~\cite{Zobelli07PRB}, electronic excitations~\cite{Banhart99RPP} and chemical etching~\cite{Meyer12PRL}. The available microscope parameters are typically limited to electron energy~\cite{Meyer12PRL}, dose rate~\cite{Robertson12NC} and temperature~\cite{Urita05PRL}. However, without control over the composition of residual gases in the microscope vacuum, there is always at least one unknown parameter in the experiments and true control cannot be achieved. Understanding and regulating the environment of the sample during electron beam exposure not only leads to improved knowledge of the interplay of the direct influence of the electron beam on the sample structure and any irradiation-induced chemical processes, but potentially can also allow one to chemically modify the material with the ultimate spatial resolution of just a few atoms.

The knock-on process is well understood, especially in graphene~\cite{Meyer12PRL,Susi16NC}, whereas electronic excitations do not seem to play a role in this material~\cite{Susi16NC}. Since knock-on damage in graphene can be effectively prevented by lowering the acceleration voltage to below 80~kV, chemical etching remains as the only damage mechanism at low voltages. In contrast to knock-on damage, chemical etching of thin materials has not been systematically studied at atomic resolution under a well-defined gas atmosphere. Due to the common design choice of side entry holders, vacuums do not typically reach pressures below $10^{-7}$~mbar, providing a large number of molecules that can participate in the etching process. Although environmental TEM in principle provides a controlled atmosphere, the typical pressures are relatively high, ranging from $10^{-2}$~mbar to tens of mbar~\cite{jinschek_advances_2014}. Such pressures are useful for many studies ranging from the growth of materials to gas sensing, but do not allow atomic-scale study of etching processes. While electron-beam-induced etching of graphene has been studied in a scanning electron microscope~\cite{Thiele13C}, the details of the process could not be elucidated due to the lack of atomic resolution. For carbon nanotubes it has been demonstrated that chemical etching is responsible for irradiation damage below the knock-on threshold by cutting them with the electron beam under different atmospheres~\cite{yuzvinsky_precision_2005,molhave_electron_2007}. In these studies, water molecules were identified as the most effective agent for electron-irradiation-induced chemical etching~\cite{yuzvinsky_precision_2005}. Water vapor has also been used to successfully etch nanoscale features in suspended graphene in a scanning electron microscope~\cite{sommer_electron-beam_2015}.

In this study, we use the customized Vienna Nion UltraSTEM 100 with a base pressure of $10^{-10}$~mbar (Methods) to mimic a typical (S)TEM device by introducing small amounts of air into the column and study its effect on the observation of the ubiquitous hydrocarbon contamination on graphene. We find that electron irradiation with some air is effective in cleaning graphene from contamination, but completely ineffective in damaging the pristine lattice. However, when defects are encountered from underneath the receding contamination, pores start to grow in graphene. The same effect can be obtained through the introduction of O$_2$ or H$_2$O into the column, whereas H$_2$ and N$_2$ have no influence as compared to experiments at $10^{-10}$~mbar, showing that chemical etching of carbonaceous materials does not necessitate introduction of water molecules.

\section{\label{sec:results}Results}

Electron-beam-induced chemical etching is fundamentally different from knock-on damage as it is not directly caused by the electrons, but by radicals that are created from residual gas molecules that are in the microscope vacuum when those get adsorbed on the sample surface~\cite{Lobo08NT,Randolph11APL}. For modeling beam-induced etching processes, the adsorption of the gas onto the surface, desorption from it, dissociation and diffusion on the surface all have to be considered~\cite{Lassiter08NT}. Here we limit the study to experimental observations of the influence of the gas atmosphere on the sample and its contamination.

\begin{figure}
\includegraphics[width=\linewidth]{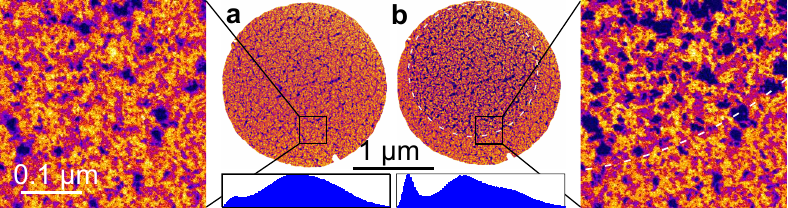}
    \caption{Influence of residual gas on graphene contamination under broad beam illumination. (a) Suspended graphene covered by carbon-based contamination (a STEM-MAADF image in UHV). The inset shows the histogram of gray values in the image (excluding the white pixels corresponding to the sample support). The darkest areas correspond to clean graphene, and the thickness of the contamination increases with increasing intensity. (b) Same sample area imaged after the sample was exposed to air at a pressure of $\sim 2\times 10^{-6}$~mbar while being homogeneously irradiated within the area shown with the dashed white circle for 32~min (ca. $1.9\times 10^{11}~$e$^-/\mu\text{m}^2$). Inset shows a clear difference in the intensities as compared to panel a.}
\label{fig:ccd_cleaning}
\end{figure}

\begin{figure}
\includegraphics[width=\linewidth]{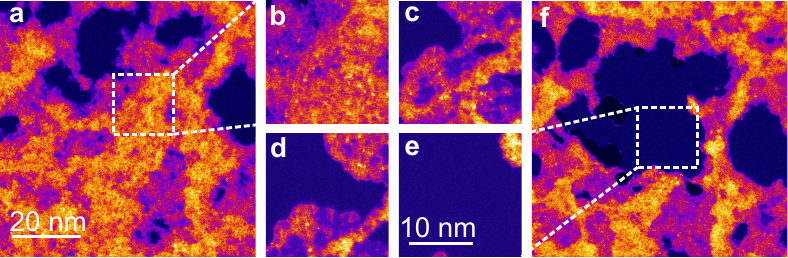}
\caption{STEM-MAADF images recorded over 22~min during continuous imaging. The dose rate was $\sim2\times 10^{8}~\text{e}^-/\text{s}$ and the pressure $\sim 1\times 10^{-6}$~mbar (air). (a) Overview image at the beginning of the experiment. (b) First image of the sequence. (c) Image recorded after an accumulated electron dose of $\sim 2\times 10^{8}~\text{e}^{-}/\text{nm}^2$. (d) Image recorded after an accumulated dose of $\sim 4\times 10^{8}~\text{e}^{-}/\text{nm}^2$. (e) Last image of the sequence after a cumulative dose of $\sim 7\times 10^{8}~\text{e}^{-}/\text{nm}^2$, showing mostly clean graphene. (f) Overview image recorded after the experiment. The regions with the the darkest contrast are pores.}
\label{fig:maadf_series}
\end{figure}

Initial experiments were conducted to determine the impact of air on the structure of sample contamination to mimic experiments carried out in an instrument with a conventional side entry design. Although the actual composition of the residual gas changes over time due to different pumping efficiencies of the constituent gases in air, and also depends on the diffusion of the gases through the O-ring, this provides a reasonable model. After these experiments the influence of each individual gas (N$_2$, O$_2$, H$_2$O, H$_2$) was measured separately. First, reference series were recorded at $2\times 10^{-9}$~mbar, i.e., in ultra high vacuum (UHV). Then the pressure in the microscope was increased with air to $\sim 1\times 10^{-6}$~mbar. First, the effect of air on the specimen was observed through the Ronchigram camera at low magnification. Fig.~\ref{fig:ccd_cleaning} shows the impact of air under parallel beam illumination at low magnification. In UHV the contamination does not change much, but in high vacuum (HV) the contaminated area starts to shrink and the contamination is concentrated into a smaller area (like reported in Ref.~\citenum{Pantelic12SSC}). Next, medium angle angular dark field (MAADF) image series were recorded in UHV and HV ($\sim 1\times 10^{-6}$~mbar, Fig.~\ref{fig:maadf_series}). These images show a similar shrinking and condensation of contamination, although it occurs now more rapidly. This is due to a much higher electron flux that causes a significantly higher concentration of radicals.

To obtain quantitative understanding on the rate of etching of the contamination (or equivalently, the rate of cleaning of graphene), we calculated the intensity of the contamination $I_{\text{c}}$ from each image recorded during an experiment similar to that shown in Fig.~\ref{fig:maadf_series}. We first removed a vacuum reference intensity from the images (value corresponding to imaging a hole), and then defined the contamination intensity as the intensity of the image $I$ with the contribution of clean graphene $I_{\text{g}}$ subtracted, normalized to $I_{\text{g}}$:

\begin{equation}
I_{\text{c}}(\phi)=\left(I(\phi)-I_{\text{g}}\right)/I_\text{g},
\end{equation}
where $\phi$ is the electron dose per area. Vacuum and graphene intensities were estimated for each image sequence by measuring intensities corresponding to a clean graphene area and an area where a one-atom-thick contamination layer covered graphene. When possible, intensity corresponding to vacuum was measured separately. This way, $I_\text{c}=1$ corresponds roughly to the intensity of a single layer of carbon atoms, allowing an intuitive understanding of the amount of contamination.

\begin{figure}
\includegraphics[width=\linewidth]{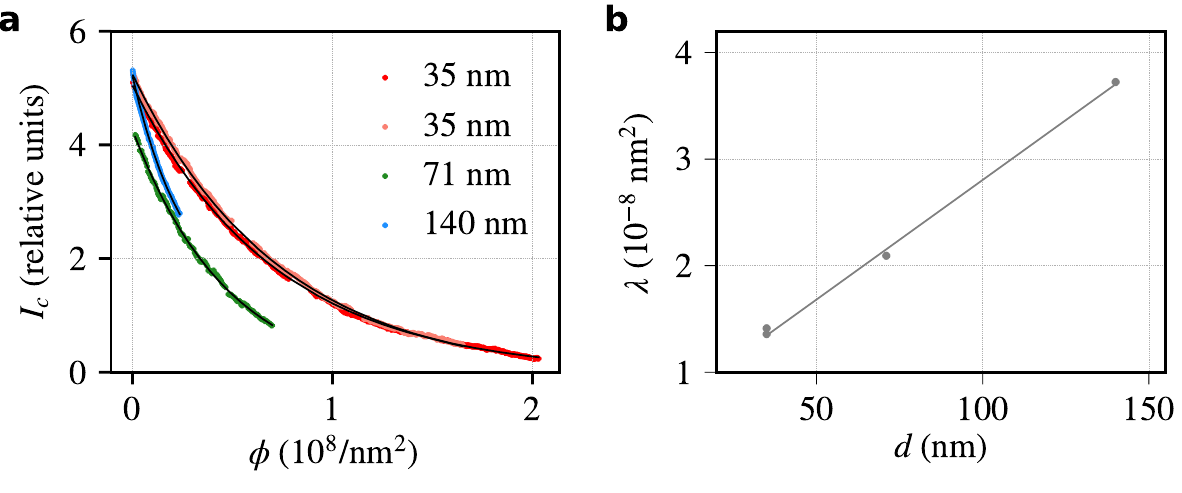}
    \caption{Etching of contamination as a function of the electron dose $\phi$ at a pressure of $\sim 2~\times 10^{-6}$~mbar (air). (a) Contamination intensity $I_{\text{c}}$ as a function of the electron dose $\phi.$ Solid lines are fits of Eq.~\ref{eq:IcExp} to the data. (b)~Decay constant $\lambda$ as a function of the field of view $d$. The solid line is provided as a guide to the eye. Uncertainties of the fits (variance of the parameter estimate) are contained within the markers. The number of pixels in the 35-nm image sequences was $512\times 512$ and for the others $1024\times 1024$.}
\label{fig:dose-effect}
\end{figure}

The data for air is shown in Fig.~\ref{fig:dose-effect}a for three different fields of view $d$ (35, 71 and 140 nm; the total square imaged area corresponds to $d^2$). The intensity data was only measured until first holes appeared in graphene (discussed below) to limit the analysis to etching of the contamination. As is clear from the image, the contamination intensity can be described as an exponential decay, showing that the amount of contamination decreases always by the same fraction for a fixed dose, regardless at what point of experiment this is measured. Hence, the intensity can be calculated via
\begin{equation}
I_\text{c}(\phi) = I_{\text{c}}(\phi=0)\exp(-\lambda\phi),
\label{eq:IcExp}
\end{equation}
where $\lambda$ is a decay factor. Thus, $\ln(2)/\lambda$ corresponds to the electron dose that is required to etch away half of the contamination and can be considered a {\it critical dose} related to the etching process. Fits of Eq.~\ref{eq:IcExp} to the data are shown as solid lines in Fig.~\ref{fig:dose-effect}a. Interestingly, it is apparent from the data that $\lambda$ depends on $d$. This relationship is shown in Fig.~\ref{fig:dose-effect}b. It appears linear, which can be interpreted to mean that the number of active etching sites in the structure scales linearly with $d$. A possible explanation for this observation is that the etching takes place at the edges of the contamination surrounding clean graphene areas (the length of those edges also increases linearly with $d$). This is also supported by the fact that the thickness of the contamination does not visually change during the experiment. Instead, the clean graphene areas appear to grow by increasing their circumference over time. Hence, we hypothesize that the electron beam cracks molecules that have landed on clean graphene, which then migrate to the contamination and participate in the etching process at the active etching sites. We point out that in order for this mechanism to be active, some clean areas must be within the field of view (notice that the contamination in the images is both sides of graphene, and one side can be clean even if the other side is covered), and thus the linear relationship observed in Fig.~\ref{fig:dose-effect}b may not extend beyond the parameter range of our experiments, especially towards the smallest possible values of $d$. However, further experiments at various different fields of view and dose rates are required to either confirm or disprove this hypothesis.

\begin{figure}
\includegraphics[width=\linewidth]{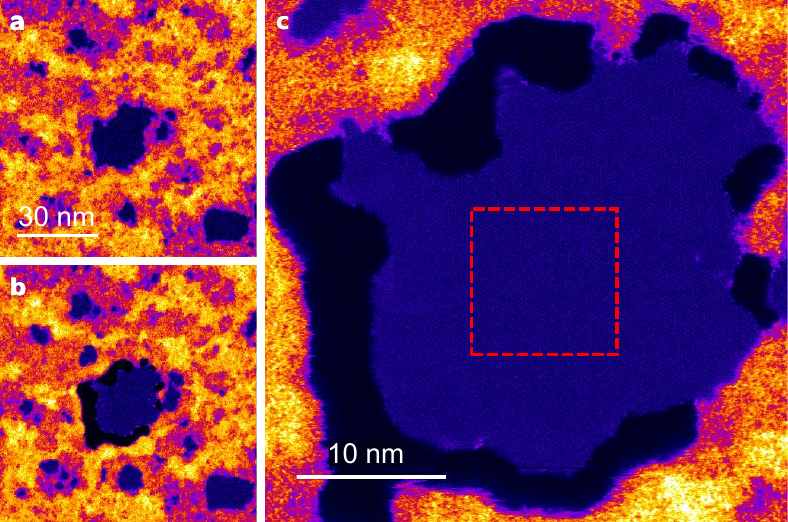}
    \caption{Chemical etching of graphene, initiated at defects in the lattice that were uncovered by receding contamination. (a) STEM-MAADF image of a small defect-free patch of graphene at the center of the frame that was exposed to electron irradiation for over 10~min (ca. $1.2\times 10^{9}$~e$^-/\text{nm}^2$) at a pressure of $2\times 10^{-6}$~mbar of air. (b) After irradiation, areas outside the scanned field of view were exposed from under the contamination and significantly damaged, with some cleaning evident also beyond the clean patch. (c) Higher magnification image of the middle of the area shown in panel b. No defects were observed in the pristine lattice despite a high cumulative dose. The red dashed square indicates the irradiated area. The regions with darkest contrast are pores.}
\label{fig:pristine-defect}
\end{figure}

No defects are created in graphene regions that were pristine at the beginning of imaging. However, the receding contamination can reveal defects in the graphene lattice, which in turn can serve as a starting point for chemical etching of the material itself, as seen in Fig.~\ref{fig:pristine-defect}. In this experiment, a small clean graphene area without any contamination within the field of view was exposed to electron irradiation for 10~min (ca. $1.2\times 10^{9}$~e$^-/\text{nm}^2$) at a pressure of $10^{-6}$~mbar of air. Throughout the experiment, no effect was seen in the images recorded of the clean graphene patch. After the experiment, in the subsequently recorded overview images, it was obvious that during the experiment moderate cleaning of the surrounding area had occurred. More interestingly, the irradiated clean graphene area was now surrounded by pores that had grown outside the field of view (the dark areas in Figs.~\ref{fig:pristine-defect}b,c). We point out that the pores are more pronounced at the left side of the image, most likely due to the longer time the electron beam spends there during a scan. While it is not directly clear from this data, the pores presumably were etched at positions where defects were uncovered by the receding contamination.

\begin{figure}
\includegraphics[width=\linewidth]{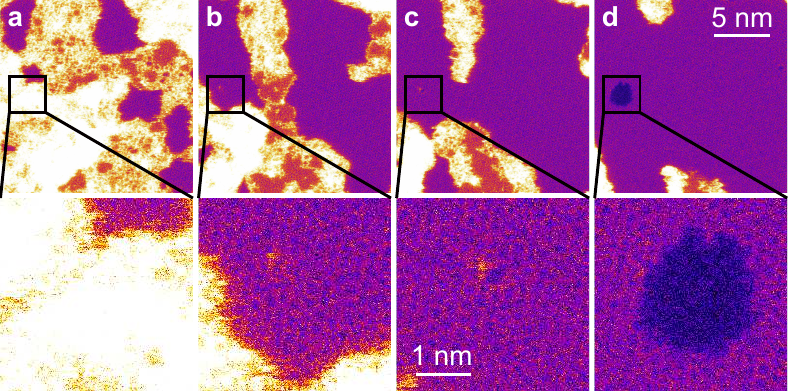}
    \caption{Creation of a nanopore starting from a point defect that is initially covered by contamination. (a)~STEM-MAADF image of graphene covered by contamination at the beginning of the experiment. (b)~A heavier atom incorporated into the graphene lattice, uncovered by the receding contamination during imaging in air ($2\times 10^{-6}$~mbar). (c-d)~A small pore starts to grow at the site of the heavier atom. The bottom row shows magnified views of the indicated areas. Total imaging time was approximately 5~min (ca. $2\times 10^{8}$~e$^-/\text{nm}^2$). The region with darkest contrast is a pore.}
\label{fig:point-defect}
\end{figure}

In another experiment (Fig.~\ref{fig:point-defect}), a defect involving an impurity atom (most likely silicon~\cite{zhou_direct_2012,ramasse_probing_2013}) was uncovered during the experiment. During continuous electron irradiation, a pore first appears and later grows at this exact location. In Ref.~\citenum{borrnert_amorphous_2012}, the authors saw a very similar appearance of pores in graphene from under receding contamination. They interpreted the formation of pores to be caused by heating of the small amorphous carbon patches on top of graphene due to the intense irradiation in STEM. If this were true, we would also see this happening in our experiments in UHV---which we do not. Hence, chemical etching initiated at defects uncovered by the receding contamination remains as the only plausible explanation.

Since the experiments with air leave the identity of the active molecular species unclear, we carried out additional experiments with other gases, including N$_2$, O$_2$, H$_2$O vapor and H$_2$. Unsurprisingly, and in agreement with the experiments by Yuzvinsky et al.~\cite{yuzvinsky_precision_2005}, H$_2$ and N$_2$ had no effect as compared to UHV ($8\times 10^{-10}$~mbar) conditions. As expected, H$_2$O had a clear effect similar to air. However, also O$_2$ lead to etching very similar to air, showing that oxygen is sufficient for the etching process.

\begin{figure*}
\includegraphics[width=\linewidth]{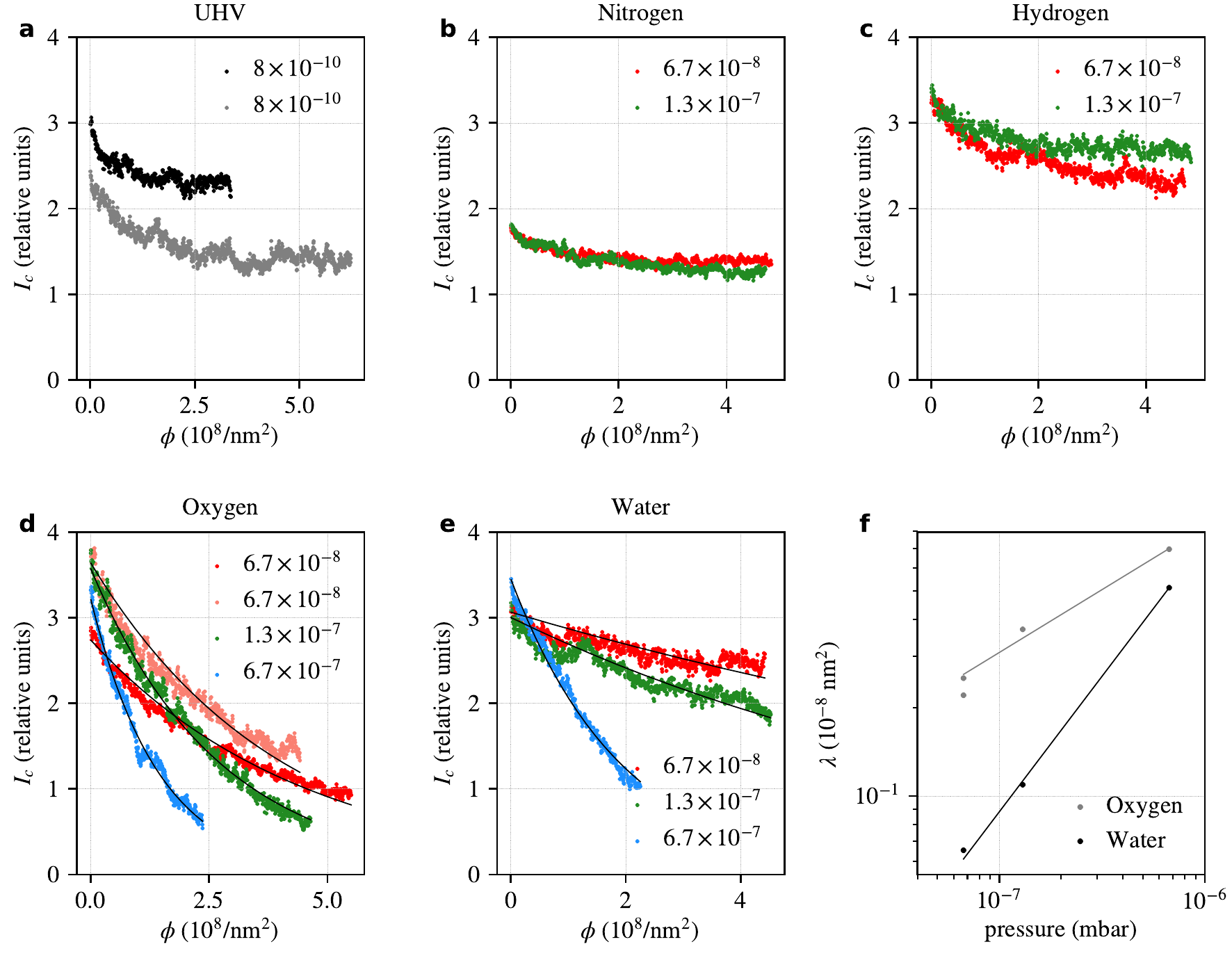}
\caption{Etching of contamination with different gases. Contamination intensity $I_\text{c}$ as measured (a) at UHV ($8\times 10^{-10}$~mbar) and (b) in N$_2$, (c) H$_2$, (d) O$_2$ and (e) H$_2$O atmospheres as a function of the electron dose $\phi$. Solid lines are fits to the data using Eq.~\ref{eq:IcExp}. (f) Decay constant $\lambda$ for O$_2$ and H$_2$O as a function of pressure (notice the logarithmic axes). The solid lines are provided as guides to the eye. Uncertainties of the original fits (variance of the parameter estimate) are contained within the markers (note that they do not necessarily reflect the match between the equation and the data).}

\label{fig:gases}
\end{figure*}

The data for UHV and all studied gases and water vapor are shown in Fig.~\ref{fig:gases}. The contamination intensity $I_\text{c}$ was measured similarly to the case of air, presented above. As can be seen from the data, in all cases there is an initial short period where some decrease in the contamination intensity is observed. We assume this to be due to the oxygen atoms present in the contamination until they are removed via the etching mechanism due to the exposure to the electron beam. After this, the values for UHV, N$_2$ and H$_2$ quickly level off to constant values (Fig.~\ref{fig:gases}a-c). These data can not be described via Eq.~\ref{eq:IcExp} since they do not tend towards zero contamination. For N$_2$ and H$_2$, increasing the pressure up to $5\times 10^{-7}$~mbar started to have a measurable effect on the contamination, but still too slow to be described by the exponential function. This is probably due to trace amounts of water or oxygen either in the gases or in the gas line. In contrast, with water and O$_2$ the etching process is similar to what we observed with air. For each case, we carried out measurements at three different pressures, shown in Fig.~\ref{fig:gases}d,e. As can be seen from the plots, the behavior is clearly exponential, but for a currently unknown reason is not described by Eq.~\ref{eq:IcExp} as well as the data recorded for air. For the highest pressures, the analysis of the sequences was terminated when holes appeared in graphene, similar to the data for air.

For O$_2$ and H$_2$O the lowest pressure data agrees more poorly with Eq.~\ref{eq:IcExp} than the data from higher pressures, as is obvious from the fits in Fig.~\ref{fig:gases}d,e. As can be seen in Fig.~\ref{fig:gases}f, the decay constant $\lambda$ depends linearly on the order of magnitude of the pressure. It is also obvious that for the same pressure, O$_2$ leads to roughly two times more efficient etching than H$_2$O, which we attribute to the two oxygen atoms per molecule in O$_2$ as compared to only one in H$_2$O. We point out that although the logarithmic relationship shown in Fig.~\ref{fig:gases}f appears to describe our data extremely well, it is not clear whether it would extend beyond our experimental range of pressures. As mentioned above, especially at lower pressures the etching becomes inefficient and can not be described with Eq.~\ref{eq:IcExp} anymore.

\section{\label{sec:conclusions}Discussion and conclusions}
Our results show the effect of residual gas on the atomic structure of the carbon-based contamination on graphene during STEM imaging. While the contamination remains practically unchanged when imaged in UHV, small amounts of air lead to clearly observable chemical etching. Pristine areas of graphene remain unperturbed by etching due to the chemically inert nature of the perfect lattice. Catalytic growth of pores in graphene has been discussed for different metal adatoms~\cite{ramasse_direct_2012}. However, our results indicate that also defect sites can activate the etching process in graphene and lead to the formation and enlargement of pores during continuous imaging. To understand in detail the influence of different defects and impurity atoms in the effectiveness of the etching process, further detailed studies are required. Here, to estimate the role of the different constituent gases in air on the etching process, we measured the etching rate in different gaseous atmospheres. In contrast to air, H$_2$O and O$_2$, experiments in H$_2$ and N$_2$ atmospheres lead to no changes in the contamination, resembling experiments carried out in UHV. The strongest etching effect was observed for oxygen, showing that O radicals are responsible for etching. The etching efficiency increases linearly with the order of magnitude of the pressure in the column as well as the field of view. The latter indicates that the etching takes place at active sites on contamination edges surrounding clean graphene areas. In conclusion, these experiments demonstrate the possibility to carry out spatially localized chemical experiments {\it in situ} in the transmission electron microscope at very low pressures, between $10^{-10}$ and $10^{-6}$~mbar, while maintaining atomic resolution imaging conditions. This work opens the way for many applications ranging from controlled cleaning of samples~\cite{tripathi_cleaning_2017} to manufacturing nanostructures with atomic precision through accurate control of electron-beam-induced chemical processes.

\section{\label{sec:system}Methods}
All STEM experiments were carried out using a modified Nion UltraSTEM~100 in Vienna. Modifications include a remodelled objective area providing several additional ports (most having line-of-sight access to the sample~\cite{hotz_modified-nion-column_2016}), a titanium-sublimation pump as well as a non-evaporable getter pump. There is an all-metal variable leak valve mounted on a port near the sample area allowing for flow rates as low as $10^{-10}$~mbar~l/s. The samples were commercial graphene grown via chemical vapor deposition and suspended on Quantifoil{\textregistered} TEM grids by the supplier Graphenea. Samples were baked in vacuum overnight at 150$^\circ$C before being inserted into the microscope. After the vacuum bake, the samples were inserted into the ultra-high vacuum system of the microscope. This procedure takes less than half an hour. The microscope was operated at 60~kV and images were recorded using the medium-angle annular dark field (MAADF) detector. In these experiments, a base pressure of $8\times 10^{-10}$~mbar was reached (the current base pressure of the device is around $2\times 10^{-10}$~mbar). By carefully opening the leak valve while monitoring the pressure, pressure levels up to $2\times 10^{-6}$~mbar could be obtained while continuous imaging at atomic resolution. In the dynamic equilibrium, reached a few seconds after adjusting the leak valve, the leak rate equals the absorption in the ion getter pump of the objective area. The maximum leak rate, and hence the maximum pressure in the objective area, is determined by the tolerable load on the ion getter pump. No change in pressure was observed in the column volumes above and below the sample area, indicating that the small apertures along the column provide excellent vacuum isolation. Initial experiments were carried out using ambient air. Later, pure gases (nitrogen, oxygen, and hydrogen) and water vapor were connected to the leak valve via a gas transfer line, which was flushed with argon and pumped to vacuum multiple times between gases.

\begin{figure*}
\includegraphics[width=\linewidth]{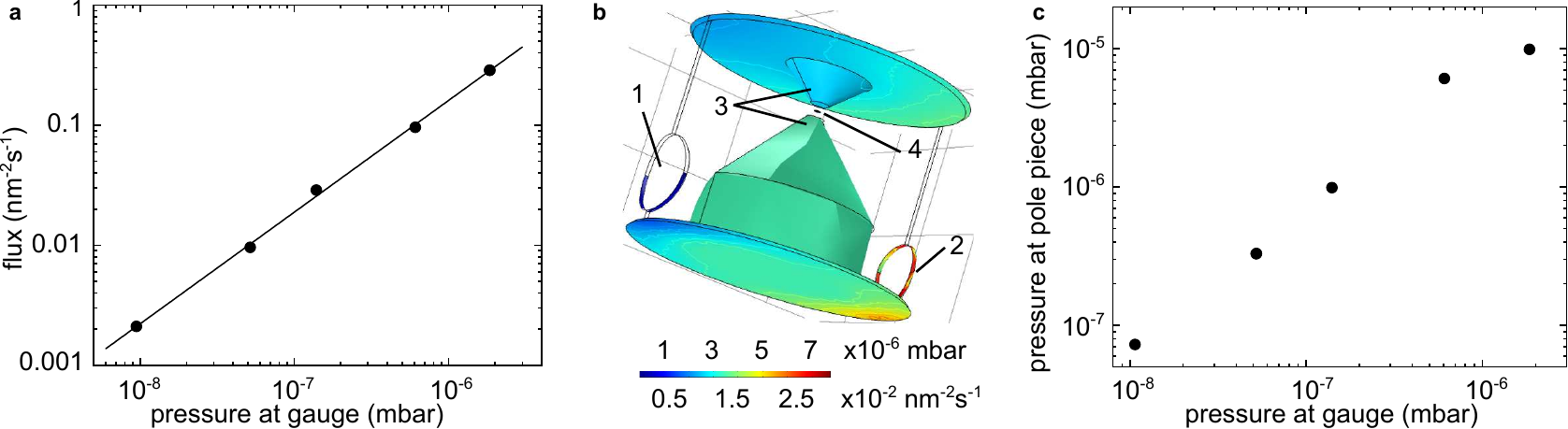}
\caption{(a) Flux at the sample as a function of the pressure at the pressure gauge close to the objective lens. The linear fit is a guide to the eye. (b) Geometry used for simulating the flux at the sample. The color scheme shows the flux (and the corresponding pressure) at different areas of the objective area. The distance between inlet and outlet is ca. 14.6~cm. The labels mark the locations of (1) the pressure gauge, (2) gas inlet, (3) pole pieces and (4) the sample. (c) The pressure at the pole piece as a function of the displayed pressure at the gauge.}
\label{fig:simulatedpressure}
\end{figure*}

The gas flow within the column depends on the geometry of the sample region inside the microscope. We therefore established a finite element model based on a simplified computer aided design schematic of the sample area. Since inlet gas flow is not directly measured, we simulated a molecular flow model with the following boundary conditions: (1) pressure at the sample is set to the initial pressure at the gauge close to the objective, (2) ion getter pumping rate is set to 1 (all incoming gas molecules are removed) and (3) no adsorption at the inner walls of the chamber (adsorption and desorption rate are identical in equilibrium). We then varied the inlet flow until the outlet pressure reached the pressure level of the experiment. Due to a limited knowledge of the proprietary sample stage design, the gas flow rate at the sample has been estimated by using the flow rate calculated at the very end of the magnetic pole piece. A typical pressure distribution inside the chamber for a given inlet flow is shown in Fig.~\ref{fig:simulatedpressure}. Note that the estimated pressure at the sample is ca. one order of magnitude higher than that measured by the gauge (Fig.~\ref{fig:simulatedpressure}c). All pressures mentioned in this article are readings from the gauge.

\acknowledgments
G.T.L. acknowledges support from the University of Vienna through the uni:docs fellowship programme. G.T.L., S.H. and J.K. were further supported by the Austrian Science Fund (FWF) through project I3181-N36, J.K. by the Wiener Wissenschafts\mbox{-,} Forschungs- und Technologiefonds (WWTF) via project MA14-009, T.S. by FWF via project P28322-N36, J.C.M. and C.M. by the European Research Council Grant No. 336453-PICOMAT, and J.C.M. and S.H. by the FWF via project P25721-N20. 

\bibliography{References_BibDesk,bibliography_JK,bib_GTL}

\end{document}